\documentclass[a4paper,11pt,dvipsnames]{article}
\usepackage{pos}
\usepackage{color,graphicx}
\usepackage{bm}
\usepackage{hyperref}
\hypersetup{%
colorlinks=true,%
citecolor=Blue,%
filecolor=Black,%
linkcolor=Blue,%
urlcolor=Violet
}
\usepackage[activate={true,nocompatibility},final,tracking=true,kerning=true,spacing=true,factor=1100,stretch=10,shrink=10]{microtype}
\microtypecontext{spacing=nonfrench}

\newcommand{\gev}    {\:\mathrm{GeV}}

\newcommand{\gevsq}  {\:\mathrm{GeV}^2}

\newcommand{\average}[1]{\left\langle{#1}\right\rangle}

\newcommand{\eq}[1]{Eq.~(\ref{#1})}

\title{Nuclear Deep Inelastic Scattering around $x=0.3$}
\ShortTitle{Nuclear DIS around $x=0.3$}
\renewcommand{\printHeadAuthors}{S. Kulagin and R. Petti}

\author[a]{S. A. Kulagin}
\author*[b]{R. Petti}

\affiliation[a]{Institute for Nuclear Research of the Russian Academy of Sciences, 117312 Moscow, Russia}
\affiliation[b]{Department of Physics and Astronomy, University of South Carolina, Columbia, SC 29208, USA}

\emailAdd{kulagin.physics@gmail.com}
\emailAdd{roberto.petti@cern.ch}

\abstract{Deep-inelastic scattering (DIS) data on nuclei, from deuterium to lead, indicate that nuclear effects on the partonic structure of bound nucleons are remarkably small around $x=0.3$, where valence-quark distributions peak.
We present the results of a global analysis of DIS data on the cross-section ratios $\sigma^A/\sigma^{{}^2\text{H}}$ in the range $0.25 \leq x \leq 0.35$, revealing a remarkable cancellation of nuclear effects across the entire range of nuclear targets. We also discuss the interpretation of such an observation within a microscopic model of nuclear modifications of the structure functions.
}

\FullConference{The 33rd International Workshop on Deep Inelastic Scattering and Related Subjects (DIS2026)\\
Bologna, Italy\\May 4 -- 8, 2026
}

\begin{document}
\maketitle

\section{Introduction}
\label{sec:intro}
Significant modifications of the partonic structure of bound nucleons have been found by many deep-inelastic scattering experiments using high-energy muon and electron beams at CERN, SLAC, DESY, and JLab probing a wide range of nuclear targets, from deuterium to lead~\cite{EuropeanMuon:1983wih,Arnold:1983mw,BCDMS:1985dor,BCDMS:1987upi,EuropeanMuon:1992pyr,Gomez:1993ri,Amaudruz:1995tq,NewMuon:1996yuf,Ackerstaff:1999ac,Seely:2009gt,CLAS:2019vsb,Arrington:2021vuu,HallC:2022utd,JeffersonLabHallATritium:2024las}.
A precise understanding of nuclear effects on bound protons and neutrons is therefore essential for the interpretation of hard-scattering processes across a wide range of experiments, as well as for accurate determinations of parton distribution functions (PDFs).

Experimental results are typically reported as structure-function (SF) ratios for a nucleus with mass number $A$ relative to deuterium,
$R_2^A=F_2^A(x,Q^2)/F_2^{{}^2\text{H}}(x,Q^2)$,
where $x$ is the Bjorken variable and $Q^2$ is the squared four-momentum transfer. The structure functions are normalized per nucleon. For isoscalar nuclei---those with equal numbers of protons ($Z$) and neutrons ($N$)---the ratio $R_2^A$ would be unity in the absence of nuclear modifications. Consequently, deviations of $R_2^A$ from unity are interpreted as manifestations of nuclear modifications of the bound-nucleon structure functions.
A remarkable observation from global analyses of experimental data~\cite{Gomez:1993ri,Kulagin:2004ie,Kulagin:2010gd,Weinstein:2010rt} is that $R_2^A$ is very close to unity around $x\approx0.3$, where valence-quark distributions dominate, indicating a cancellation of nuclear corrections in that region. Rather small nuclear modifications are observed over the interval $0.25\leq x\leq0.35$ across a wide range of nuclear targets and $Q^2$ values.

A quantitative assessment of the cancellation of nuclear effects for $0.25\leq x\leq0.35$ was presented in a recent study~\cite{Kulagin:2026wiv} using world DIS data.
Here we summarize the main observations (Sec.~\ref{sec:data}) and compare the data with predictions of the microscopic nuclear model of Refs.~\cite{Kulagin:2004ie,Kulagin:2010gd,Kulagin:2014vsa}.
We then discuss the physical mechanisms responsible for the cancellation of nuclear modifications within such a narrow kinematic window (Sec.~\ref{sec:discus}).

\section{Data analysis}
\label{sec:data}

For non-isoscalar nuclei ($Z \neq N$), the ratio $R_2^A$ differs from unity even in the absence of nuclear modifications, simply because the proton and neutron structure functions are different. It is therefore customary to correct the measured $F_2^A$ and $R_2^A$ for non-isoscalarity according to
{\allowdisplaybreaks
\begin{align}
\label{eq:f2is}
F_2^{A(\text{is})} &= F_2^A
\frac{A F_2^N}{Z F_2^p + N F_2^n}
= F_2^A
\frac{A(1+R_{np})}{2(Z+N R_{np})},
\\
R_2^A &= F_2^{A(\text{is})}/F_2^{{}^2\text{H}},
\label{eq:r2}
\end{align}
}%
where $F_2^p$ and $F_2^n$ are the free-proton and free-neutron structure functions, respectively, and $R_{np}=F_2^n/F_2^p$. We also use the ratio
\begin{align}
\label{eq:ris}
R_2^{A(\text{is})}
&= \frac{A F_2^A}{Z F_2^p + N F_2^n}
= \frac{F_2^{A(\text{is})}}{F_2^N},
\end{align}
where
$F_2^N=\frac{1}{2}(F_2^p+F_2^n)$
is the structure function of an isoscalar free nucleon. Although not directly measurable, $R_2^{A(\text{is})}$ provides a useful measure of the magnitude of nuclear effects for a generic nucleus with mass number $A$ and charge $Z$.

In a recent study~\cite{Kulagin:2026wiv}, we presented a global analysis of $R_2^A$ data from the available DIS experiments~\cite{BCDMS:1985dor,BCDMS:1987upi,EuropeanMuon:1992pyr,Dasu:1993vk,Gomez:1993ri,Amaudruz:1995tq,NewMuon:1996yuf,Ackerstaff:1999ac,Garutti:2003,Seely:2009gt,CLAS:2019vsb,Arrington:2021vuu,HallC:2022utd,JeffersonLabHallATritium:2024las}, focusing on the kinematic region $0.25\leq x\leq0.35$. For each nuclear target, we combined the measurements within the selected $x$ interval into a weighted average $\bar R_2^A$. The averaging procedure accounted for the different types of experimental uncertainties and, when measurements of the same target from different experiments were available, combined them consistently~\cite{Kulagin:2026wiv}.

\begin{figure}[htb!]
\centering
\includegraphics[width=0.7\textwidth]{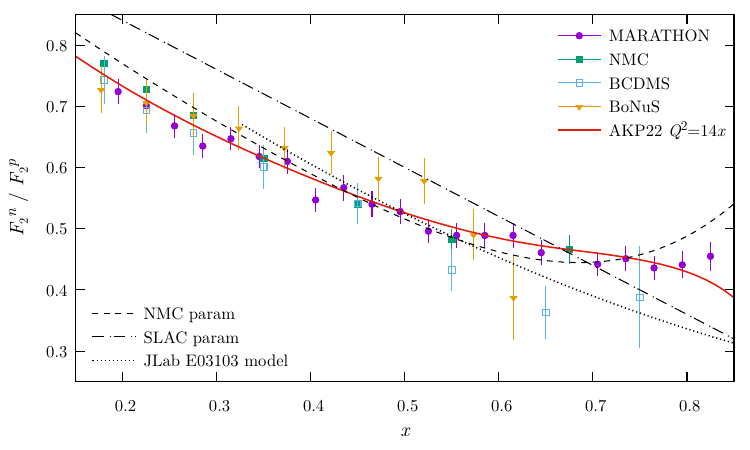}%
\caption{%
Ratio $R_{np}=F_2^n/F_2^p$ measured by the MARATHON experiment~\cite{JeffersonLabHallATritium:2021usd} compared with the measurements from NMC~\cite{NewMuon:1996uwk}, BCDMS~\cite{BCDMS:1989gtb}, and BoNuS~\cite{CLAS:2014jvt} ($W\geq1.8\gev$), all evolved to the MARATHON reference scale $Q^2=14x\gevsq$.
The solid red curve shows the prediction obtained from the proton and neutron structure functions of the AKP22 global QCD fit~\cite{Alekhin:2022uwc}.
For comparison, the phenomenological parameterizations from NMC~\cite{NewMuon:1991exl} and SLAC~\cite{Gomez:1993ri}, together with the JLab E03-103 model prediction~\cite{Arrington:2021vuu}, are shown at the same scale.
}
\label{fig:Rnp-comp}
\end{figure}

Non-isoscalarity corrections applied in published measurements often rely on inconsistent inputs for the ratio $R_{np}$ in \eq{eq:f2is}, thereby introducing a potential source of model-dependent systematic uncertainty. To assess the impact of such differences, we recalculated the correction for each data point using a common $R_{np}(x,Q^2)$ obtained from a parameterization of the MARATHON~\cite{JeffersonLabHallATritium:2021usd} and NMC~\cite{NewMuon:1991exl} measurements. 
Figure~\ref{fig:Rnp-comp} compares the recent MARATHON measurement of $R_{np}$~\cite{JeffersonLabHallATritium:2021usd} with the corresponding determinations by NMC~\cite{NewMuon:1996uwk}, BCDMS~\cite{BCDMS:1989gtb}, and BoNuS~\cite{CLAS:2014jvt}, together with the prediction from the AKP22 global QCD fit~\cite{Alekhin:2022uwc}.
Since the datasets were collected at different $Q^2$ values for a given $x$, a direct comparison requires their evolution to a common reference scale. We evolved the NMC, BCDMS, and BoNuS data to the reference scale of the MARATHON measurement, $Q_M^2=14x\gevsq$, by multiplying each $(x,Q^2)$ data point by the corresponding correction factor $R_{np}(x,Q_M^2)/R_{np}(x,Q^2)$. The measurements are in excellent mutual agreement in the region of interest, $0.2<x<0.4$. The AKP22 prediction~\cite{Alekhin:2022uwc} is also in good agreement with both the MARATHON and NMC data over the full kinematic range considered, providing a consistent description of the available measurements and a baseline for a common non-isoscalarity correction. For comparison, Fig.~\ref{fig:Rnp-comp} also shows the phenomenological parameterization obtained from the NMC fit to the NMC, BCDMS, and SLAC $R_{np}$ data~\cite{NewMuon:1991exl}, the legacy parameterization used in the SLAC E139 analysis~\cite{Gomez:1993ri}, and the JLab E03-103 model~\cite{Arrington:2021vuu}, all evolved to $Q^2=Q_M^2$.

\begin{figure}[htb]
\centering
\includegraphics[width=1.00\textwidth]{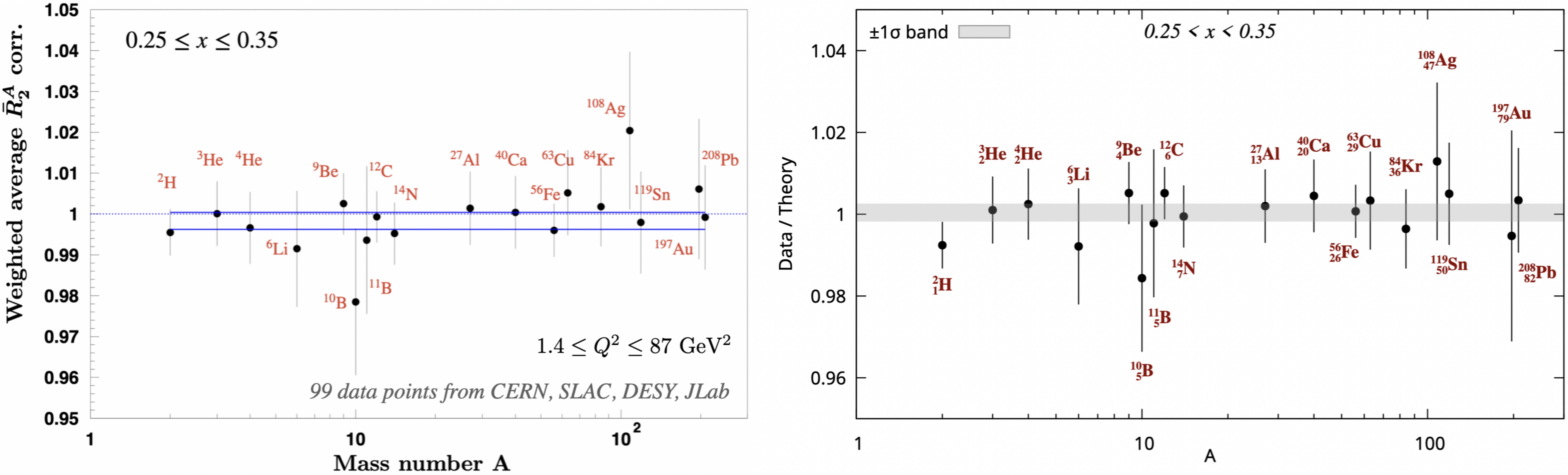}
\caption{%
(Left) Weighted averages of the $\bar R_2^A$ measurements in the range $0.25\leq x\leq0.35$, with a consistent correction for non-isoscalarity (see text).
The MARATHON measurements are not included in the weighted averages. The solid lines indicate the $\pm1\sigma$ band obtained from a constant fit to all nuclei.
(Right) Ratio of the averaged $\bar R_2^A$ measurements to the corresponding predictions of the microscopic model~\cite{Kulagin:2004ie,Kulagin:2010gd} in the interval $0.25\leq x\leq0.35$. The same weighted-averaging procedure is applied to the experimental data and theoretical predictions. The shaded area represents the $\pm1\sigma$ uncertainty band obtained from a constant fit to the full dataset.
}
\label{fig:R2exp}
\end{figure}

The resulting weighted averages $\bar R_2^{A\,\mathrm{corr.}}$ for the individual targets are shown in the left panel of Fig.~\ref{fig:R2exp} as a function of the mass number $A$. The values of $\bar R_2^{A\,\mathrm{corr.}}$ are remarkably close to unity for all nuclei considered. A fit to these data with a constant value gives
$\average{\bar R_2^{A\,\mathrm{corr.}}}=0.9988\pm0.0023$. Replacing the individual non-isoscalarity corrections by a common correction based on the procedure described above has only a marginal effect on the result. The resulting shift in the fitted constant, $\average{\bar R_2^A}$, is substantially smaller than the corresponding fit uncertainty, as shown in Table~2 of Ref.~\cite{Kulagin:2026wiv}.
Moreover, both fits are in excellent agreement with the fit restricted to isoscalar nuclei. This result indicates that the non-isoscalarity corrections do not introduce any significant bias in the results shown in Fig.~\ref{fig:R2exp} (see also Figs.~2 and~3 of Ref.~\cite{Kulagin:2026wiv}).

\section{Discussion}
\label{sec:discus}

To understand the remarkable cancellation of nuclear effects reported in Sec.~\ref{sec:data}, we calculated the predictions of the microscopic model of Refs.~\cite{Kulagin:2004ie,Kulagin:2010gd} for each data point included in the analysis of Sec.~\ref{sec:data} (see Table~1 of Ref.~\cite{Kulagin:2026wiv}).
The model incorporates several key nuclear effects: Fermi motion and binding (FMB), implemented through the convolution with the energy-momentum distribution of bound nucleons (the nuclear spectral function); off-shell (OS) modifications of the bound-nucleon structure functions; meson-exchange currents (MEC); and nuclear shadowing (NS), arising from the propagation of the hadronic component of the virtual intermediate boson through the nuclear medium.
The model has been shown to provide an excellent description of nuclear DIS data~\cite{Kulagin:2004ie,Kulagin:2010gd,Alekhin:2017fpf,JeffersonLabHallATritium:2024las}, Drell--Yan lepton-pair production~\cite{Kulagin:2014vsa}, and $W^\pm/Z$ boson production in $p+\text{Pb}$ collisions at the LHC~\cite{Ru:2016wfx}.

The theoretical predictions were averaged using the same procedure as the experimental data (Sec.~\ref{sec:data}).
The right panel of Fig.~\ref{fig:R2exp} shows the data-to-theory ratios for the same data points displayed in the left panel. Overall, we find excellent agreement between the model and the data. A fit to the ratios in Fig.~\ref{fig:R2exp} (right) with a constant value gives
$\average{\bar R_2^A/\left(\bar R_2^A\right)_{\rm model}}=1.0004\pm0.0022$.

The MARATHON data~\cite{JeffersonLabHallATritium:2024las} on the EMC effect in ${}^3\mathrm{He}$ and ${}^3\mathrm{H}$ are not included in Fig.~\ref{fig:R2exp}.
For completeness, the ratios of the weighted averages of the MARATHON measurements to the corresponding model predictions are $0.9996\pm0.0073$ for ${}^3\mathrm{H}/{}^2\mathrm{H}$ and $0.9983\pm0.0073$ for ${}^3\mathrm{He}/{}^2\mathrm{H}$.

\begin{figure}[htb!]
\centering
\includegraphics[width=1.00\textwidth]{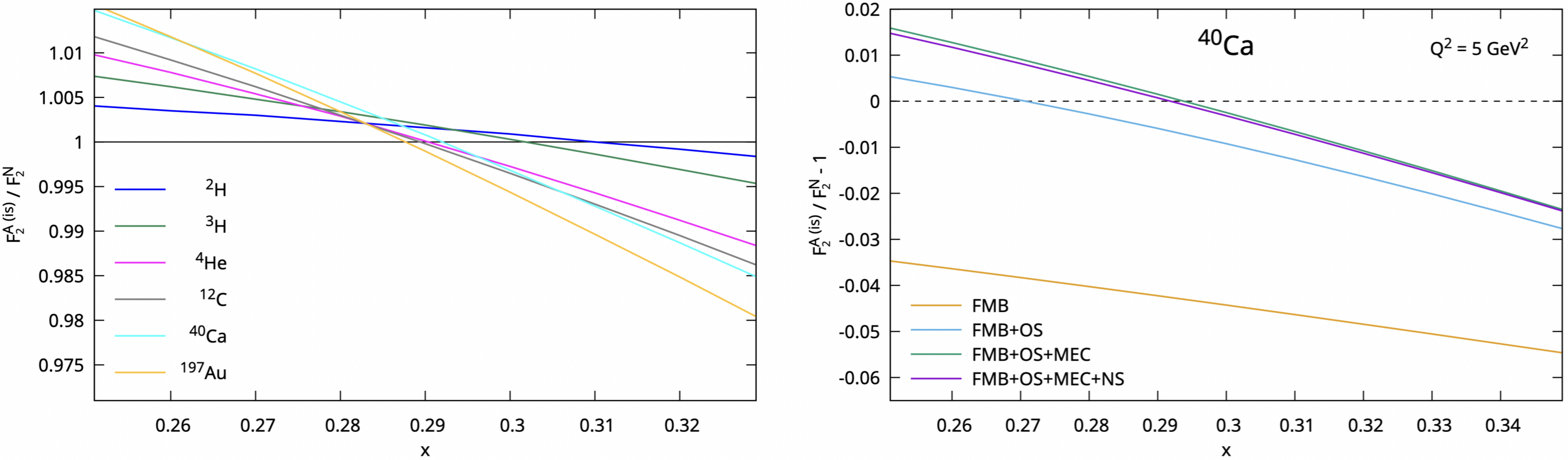}
\caption{%
(Left) Predicted ratios $R_2^{A\text{(is)}}$ for ${}^2\mathrm{H}$, ${}^3\mathrm{H}$, ${}^4\mathrm{He}$, ${}^{12}\mathrm{C}$, ${}^{40}\mathrm{Ca}$, and ${}^{197}\mathrm{Au}$ as functions of $x$ at $Q^2=5\gevsq$.
The calculations are based on the microscopic model of Refs.~\cite{Kulagin:2004ie,Kulagin:2010gd}.
(Right) Decomposition of the nuclear contributions to $R_2^{A\text{(is)}}$ for ${}^{40}\mathrm{Ca}$: nuclear smearing with the spectral function (FMB); inclusion of off-shell corrections (FMB+OS); addition of meson-exchange currents (FMB+OS+MEC); and the full result including nuclear shadowing (FMB+OS+MEC+NS).
}
\label{fig:crossover01}
\end{figure}

To investigate the physical origin of the cancellation of nuclear corrections near $x=0.3$, we calculated the ratios $R_2^{A\text{(is)}}(x,Q^2)$ for various nuclei over the interval $0.25<x<0.35$ using the microscopic model of Refs.~\cite{Kulagin:2004ie,Kulagin:2010gd}.
The results are shown in the left panel of Fig.~\ref{fig:crossover01} as functions of $x$ at $Q^2=5\gevsq$ for targets ranging from ${}^2\mathrm{H}$ to ${}^{197}\mathrm{Au}$.
Remarkably, the curves for all nuclei cross in the vicinity of a common point, where $R_2^{A\text{(is)}}\approx1.002$.
This small deviation from unity is further reduced in the experimentally measured ratio
$R_2^A=R_2^{A\text{(is)}}/R_2^{{}^2\text{H(is)}}$,
which approaches unity as a result of the normalization to the deuterium curve.

The crossover point $x_0$ is defined by the condition
\begin{equation}\label{eq:x0}
R_2^{A\text{(is)}}(x_0,Q^2)=1.
\end{equation}
In general, $x_0$ depends on both $Q^2$ and the nuclear mass number $A$.
As seen in Fig.~\ref{fig:crossover01}, at fixed $Q^2=5\gevsq$ the values of $x_0(A)$ remain within a remarkably narrow interval around $x_0=0.3$, with only a weak dependence on $A$.

\begin{figure}[htb!]
\centering
\includegraphics[width=1.00\textwidth]{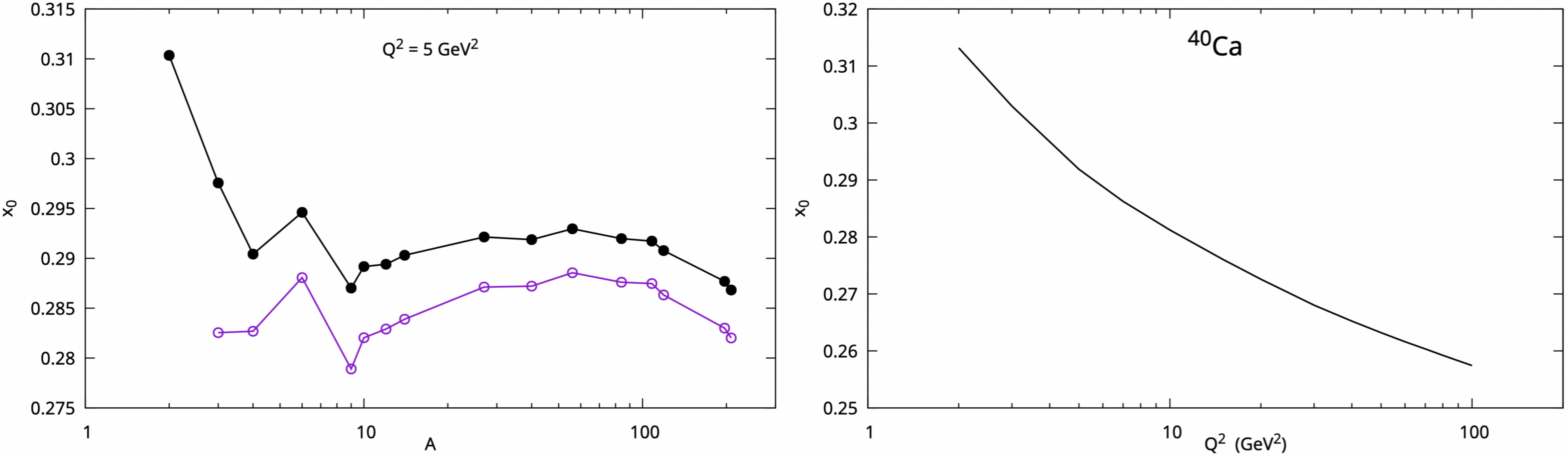}
\caption{%
(Left) The crossover point $x_0$ as a function of the mass number $A$, calculated at $Q^2=5\gevsq$ for the nuclei shown in Fig.~\ref{fig:R2exp}.
The solid symbols correspond to the solution to \eq{eq:x0}, while the open symbols represent the solution to $F_2^{A(\text{is})}(x,Q^2)=F_2^{{}^2\text{H}}(x,Q^2)$.
(Right) The $Q^2$ dependence of the crossover point $x_0$ for the ${}^{40}\mathrm{Ca}$ nucleus.
}
\label{fig:crossover02}
\end{figure}

The right panel of Fig.~\ref{fig:crossover01} shows the individual nuclear contributions for ${}^{40}\mathrm{Ca}$.
In the kinematic region of interest, $0.25\leq x\leq0.35$, the leading nuclear effects are the smearing of the proton and neutron structure functions with the nuclear energy-momentum distribution (nuclear spectral function) and the off-shell corrections to the bound-nucleon structure functions~\cite{Kulagin:2004ie}.
The spectral function employed in the model of Ref.~\cite{Kulagin:2004ie} (see also Ref.~\cite{Kulagin:2000yw}) accounts for both the mean-field component and the high-momentum tail generated by short-range nucleon-nucleon correlations.
We observe a significant cancellation between nuclear smearing and the off-shell correction in the vicinity of $x=0.3$.
The subleading contribution from nuclear meson-exchange currents therefore also plays an important role in determining the crossover point $x_0$.
Remarkably, the $A$ dependence of the MEC contribution is similar to that of the nuclear smearing and off-shell effects.
This behavior can be understood from the fact that the MEC contribution is fundamentally related to nuclear binding through sum rules~\cite{Akulinichev:1985ij,Kulagin:1989mu,Kulagin:2004ie}.
As a result, the $A$ dependence of the crossover point $x_0$ is relatively weak, as shown in the left panel of Fig.~\ref{fig:crossover02} (solid symbols).
For comparison, we also show the crossover points defined by
$F_2^{A(\text{is})}(x,Q^2)=F_2^{{}^2\text{H}}(x,Q^2)$
for the experimentally measurable nuclear ratios relative to deuterium (open symbols).
Finally, the $Q^2$ dependence of $x_0$ for ${}^{40}\mathrm{Ca}$ is illustrated in the right panel of Fig.~\ref{fig:crossover02}.

The interplay between nuclear binding and off-shell corrections in the nuclear convolution in the valence region was further discussed in Ref.~\cite{Kulagin:2026wiv} using a derivative expansion of the nuclear convolution~\cite{Akulinichev:1985ij,Kulagin:1989mu,Kulagin:1994fz}.
Let $p=(M+\varepsilon,\bm p)$ denote the four-momentum of a bound nucleon.
In the nuclear ground state, the characteristic energy $\varepsilon$ and momentum $|\bm p|$ contributing to the convolution in the region of interest are small compared with the nucleon mass $M$.
The nucleon structure function $F_2^N$ entering the convolution can therefore be expanded in powers of $\varepsilon/M$ and $\bm p/M$, leading to a derivative expansion of the nuclear convolution integral~\cite{Kulagin:1989mu,Kulagin:1994fz}.
For the ratio $R_2^{A(\text{is})}$, the leading terms in the $1/M$ expansion give
\begin{equation}
\label{eq:dexp}
R_2^{A(\text{is})}
\approx
1-e_A x\partial_x\ln F_2^N(x,Q^2)
+v_A\delta f(x),
\end{equation}
where $e_A=\langle\varepsilon\rangle/M$ is the average bound-nucleon energy and
$v_A=\langle p^2\rangle/M^2-1$ is the average nucleon virtuality.
Here $\varepsilon$ includes the nucleon separation energy $E>0$ and the recoil energy of the spectator nucleus, which balances the nucleon momentum,
$\varepsilon=-E-\frac{\bm p^2}{2M_{A-1}}$.
At leading order in $1/M$, the nucleon virtuality can be written as
$v_A=2(\langle\varepsilon\rangle-\langle T\rangle)/M$,
in terms of the average nucleon energy $\langle\varepsilon\rangle$ and the kinetic energy
$\langle T\rangle=\frac{\langle\bm p^2\rangle}{2M}$.
The averages are evaluated using the nuclear spectral function as described in Ref.~\cite{Kulagin:2004ie}.
In \eq{eq:dexp}, we neglect the second-derivative term proportional to
$\langle T\rangle x^2\partial_x^2F_2$,
as its contribution is small in the kinematic region under consideration.

The function
$\delta f=M^2\partial_{p^2}\ln F_2^N(x,Q^2,p^2)$
is the derivative of the off-shell nucleon structure function with respect to the invariant mass squared $p^2$, evaluated near the mass shell $p^2=M^2$~\cite{Kulagin:2004ie,Kulagin:2010gd}.
It characterizes the relative modification of the nucleon structure function away from the mass shell and, by definition, is universal and independent of the particular nucleus.
Off-shell modifications in nuclear DIS have been studied extensively within this framework~\cite{Kulagin:2004ie,Kulagin:1994fz,Kulagin:2010gd,Kulagin:2014vsa,Alekhin:2017fpf,Alekhin:2022tip,Alekhin:2022uwc}.
A $Q^2$-independent approximation, $\delta f(x)$, has been shown to provide an excellent description of nuclear DIS data~\cite{Kulagin:2004ie,Kulagin:2010gd,Alekhin:2017fpf,JeffersonLabHallATritium:2024las}, Drell--Yan lepton-pair production~\cite{Kulagin:2014vsa}, and $W^\pm/Z$ boson production in $p+\text{Pb}$ collisions at the LHC~\cite{Ru:2016wfx}.
Furthermore, a recent global QCD analysis~\cite{Alekhin:2022uwc} found the proton-neutron asymmetry,
$\delta f^p-\delta f^n$,
to be consistent with zero within uncertainties.
We therefore employ a single universal off-shell function $\delta f(x)$ for both protons and neutrons, as determined from the global analysis of nuclear structure functions in Ref.~\cite{Kulagin:2004ie}.

An important observation~\cite{Kulagin:2026wiv} following from \eq{eq:dexp} is that a universal, nucleus-independent crossover point $x_0$ exists if the ratio $v_A/e_A$ is independent of $A$.
This, in turn, requires the average separation and kinetic energies to have the same $A$ dependence.
Although the $A$ dependence of $\langle\varepsilon\rangle$ and $\langle T\rangle$ need not be exactly identical, their similar scaling with $A$ leads to the weak $A$ dependence of $x_0$ observed in the left panel of Fig.~\ref{fig:crossover02} (solid symbols).

The crossover point $x_0$ shifts toward lower values as $Q^2$ increases, driven primarily by the $Q^2$ evolution of the nucleon structure function $F_2^N$.
Notably, the $A$ dependence of $x_0$ is significantly weaker than its $Q^2$ dependence, as illustrated in the right panel of Fig.~\ref{fig:crossover02}.
However, existing experimental data on the $Q^2$ dependence remain sparse, both in kinematic coverage and in the range of nuclear targets.
New high-precision measurements covering a broad range of scales, $Q^2\sim1\text{--}100\gevsq$, across multiple nuclei within a single experiment would therefore be highly desirable.
To resolve the predicted shifts unambiguously, the normalization uncertainty would need to be controlled below $1\%$.
Future programs at the Electron-Ion Collider~\cite{AbdulKhalek:2021gbh} and long-baseline neutrino facilities~\cite{DUNE:2020ypp,Petti:2022bzt} could provide the sensitivity required for such measurements.

\acknowledgments
We thank S. I. Alekhin and G. G. Petratos for useful discussions.
This work is supported by Grant No.~DE-SC0026395 from the Department of Energy, USA.

\bibliographystyle{JHEP}
\bibliography{references}

@article{Kulagin:2026wiv,
    author = "Kulagin, S. A. and Petti, R.",
    title = "{On the Cancellation of Nuclear Effects in the Valence Region}",
    eprint = "2604.23110",
    archivePrefix = "arXiv",
    primaryClass = "nucl-th",
    month = "4",
    year = "2026"
}

@article{AbdulKhalek:2021gbh,
    author = "Abdul Khalek, R. and others",
    title = "{Science Requirements and Detector Concepts for the Electron-Ion Collider}: {EIC Yellow Report}",
    eprint = "2103.05419",
    archivePrefix = "arXiv",
    primaryClass = "physics.ins-det",
    reportNumber = "BNL-220990-2021-FORE, JLAB-PHY-21-3198, LA-UR-21-20953",
    doi = "10.1016/j.nuclphysa.2022.122447",
    journal = "Nucl. Phys. A",
    volume = "1026",
    pages = "122447",
    year = "2022"
}

@article{DUNE:2020ypp,
    author = "Abi, Babak and others",
    collaboration = "DUNE",
    title = "{Deep Underground Neutrino Experiment (DUNE), Far Detector Technical Design Report, Volume II: DUNE Physics}",
    eprint = "2002.03005",
    archivePrefix = "arXiv",
    primaryClass = "hep-ex",
    reportNumber = "FERMILAB-PUB-20-025-ND, FERMILAB-DESIGN-2020-02",
    doi = "10.2172/1599307",
    journal = "",
    month = "2",
    year = "2020"
}

@article{Weinstein:2010rt,
    author = "Weinstein, L. B. and others",
    title = "{Short Range Correlations and the EMC Effect}",
    eprint = "1009.5666",
    archivePrefix = "arXiv",
    primaryClass = "hep-ph",
    reportNumber = "JLAB-PHY-10-1227",
    doi = "10.1103/PhysRevLett.106.052301",
    journal = "Phys. Rev. Lett.",
    volume = "106",
    pages = "052301",
    year = "2011"
}

@article{EuropeanMuon:1983wih,
    author = "Aubert, J. J. and others",
    collaboration = "European Muon",
    title = "{The ratio of the nucleon structure functions $F_{2}^N$ for iron and deuterium}",
    reportNumber = "CERN-EP/83-14",
    doi = "10.1016/0370-2693(83)90437-9",
    journal = "Phys. Lett. B",
    volume = "123",
    pages = "275--278",
    year = "1983"
}

@article{Arnold:1983mw,
    author = "Arnold, R. G. and others",
    title = "{Measurements of the A-dependence of deep inelastic electron scattering from nuclei}",
    reportNumber = "SLAC-PUB-3257",
    doi = "10.1103/PhysRevLett.52.727",
    journal = "Phys. Rev. Lett.",
    volume = "52",
    pages = "727",
    year = "1984"
}

@article{Seely:2009gt,
    author = "Seely, J. and others",
    title = "{New measurements of the EMC effect in very light nuclei}",
    eprint = "0904.4448",
    archivePrefix = "arXiv",
    primaryClass = "nucl-ex",
    reportNumber = "JLAB-PHY-09-956",
    doi = "10.1103/PhysRevLett.103.202301",
    journal = "Phys. Rev. Lett.",
    volume = "103",
    pages = "202301",
    year = "2009"
}

@article{Gomez:1993ri,
    author = "Gomez, J. and others",
    title = "{Measurement of the A-dependence of deep inelastic electron scattering}",
    reportNumber = "SLAC-PUB-5813",
    doi = "10.1103/PhysRevD.49.4348",
    journal = "Phys. Rev. D",
    volume = "49",
    pages = "4348--4372",
    year = "1994"
}

@article{Amaudruz:1995tq,
    author = "Amaudruz, P. and others",
    collaboration = "New Muon",
    title = "{A Reevaluation of the nuclear structure function ratios for D, He, Li-6, C and Ca}",
    eprint = "hep-ph/9503291",
    archivePrefix = "arXiv",
    doi = "10.1016/0550-3213(94)00023-9",
    journal = "Nucl. Phys. B",
    volume = "441",
    pages = "3--11",
    year = "1995"
}

@article{Ackerstaff:1999ac,
    author = "Airapetian, A. and others",
    collaboration = "HERMES",
    title = "{Nuclear effects on R = $\sigma_L$ / $\sigma_T$ in deep inelastic scattering}",
    eprint = "arXiv:hep-ex/0210067",
    archivePrefix = "arXiv",
    reportNumber = "DESY-02-092",
    journal = "Phys. Lett. B",
    volume = "475",
    pages = "339--346",
    year = "2003",
}

@article{HallC:2022utd,
    author = "Karki, A. and others",
    collaboration = "Hall C",
    title = "{First Measurement of the EMC effect in B10 and B11}",
    eprint = "2207.03850",
    archivePrefix = "arXiv",
    primaryClass = "nucl-ex",
    reportNumber = "JLAB-PHY-22-3648",
    doi = "10.1103/PhysRevC.108.035201",
    journal = "Phys. Rev. C",
    volume = "108",
    number = "3",
    pages = "035201",
    year = "2023"
}

@article{CLAS:2019vsb,
    author = "Schmookler, B. and others",
    collaboration = "CLAS",
    title = "{Modified structure of protons and neutrons in correlated pairs}",
    eprint = "2004.12065",
    archivePrefix = "arXiv",
    primaryClass = "nucl-ex",
    doi = "10.1038/s41586-019-0925-9",
    journal = "Nature",
    volume = "566",
    number = "7744",
    pages = "354--358",
    year = "2019"
}

@article{NewMuon:1996yuf,
    author = "Arneodo, M. and others",
    collaboration = "New Muon",
    title = "{The A dependence of the nuclear structure function ratios}",
    reportNumber = "PRINT-97-207",
    doi = "10.1016/S0550-3213(96)90117-0",
    journal = "Nucl. Phys. B",
    volume = "481",
    pages = "3--22",
    year = "1996"
}

@article{EuropeanMuon:1992pyr,
    author = "Ashman, J. and others",
    collaboration = "European Muon",
    title = "{A Measurement of the ratio of the nucleon structure function in copper and deuterium}",
    reportNumber = "CERN-PPE-92-155",
    doi = "10.1007/BF01565050",
    journal = "Z. Phys. C",
    volume = "57",
    pages = "211--218",
    year = "1993"
}

@phdthesis{Garutti:2003,
      author         = "Garutti, E.",
      title          = "{Nuclear effects in semi-inclusieve deep-inelastic scattering off ${}^{84}$Kr and other nuclei}",
      school         = "University of Amsterdam",
      year           = "2003",
      SLACcitation   = "%%CITATION = FERMILAB-THESIS-2016-03;%%"
}

@article{BCDMS:1985dor,
    author = "Bari, G. and others",
    collaboration = "BCDMS",
    title = "{A Measurement of Nuclear Effects in Deep Inelastic Muon Scattering on Deuterium, Nitrogen and Iron Targets}",
    reportNumber = "CERN-EP-85-132",
    doi = "10.1016/0370-2693(85)90238-2",
    journal = "Phys. Lett. B",
    volume = "163",
    pages = "282",
    year = "1985"
}

@article{BCDMS:1987upi,
    author = "Benvenuti, A. C. and others",
    collaboration = "BCDMS",
    title = "{Nuclear Effects in Deep Inelastic Muon Scattering on Deuterium and Iron Targets}",
    reportNumber = "CERN-EP-87-13",
    doi = "10.1016/0370-2693(87)90664-2",
    journal = "Phys. Lett. B",
    volume = "189",
    pages = "483--487",
    year = "1987"
}

@article{Dasu:1993vk,
    author = "Dasu, S. and others",
    title = "{Measurement of kinematic and nuclear dependence of $R = \sigma_L / \sigma_T$ in deep inelastic electron scattering}",
    reportNumber = "SLAC-PUB-5814, UR-1304, ER-40685-753",
    doi = "10.1103/PhysRevD.49.5641",
    journal = "Phys. Rev. D",
    volume = "49",
    pages = "5641--5670",
    year = "1994"
}

@article{JeffersonLabHallATritium:2021usd,
    author = "Abrams, D. and others",
    collaboration = "Jefferson Lab Hall A Tritium",
    title = "{Measurement of the Nucleon $F^n_2/F^p_2$ Structure Function Ratio by the Jefferson Lab MARATHON Tritium/Helium-3 Deep Inelastic Scattering Experiment}",
    eprint = "2104.05850",
    archivePrefix = "arXiv",
    primaryClass = "hep-ex",
    reportNumber = "JLAB-PHY-21-3356",
    doi = "10.1103/PhysRevLett.128.132003",
    journal = "Phys. Rev. Lett.",
    volume = "128",
    number = "13",
    pages = "132003",
    year = "2022"
}

@article{NewMuon:1991exl,
    author = "Amaudruz, P. and others",
    collaboration = "New Muon",
    title = "{The ratio $F_2^n / F_2^p$ in deep inelastic muon scattering}",
    reportNumber = "CERN-PPE-91-167",
    doi = "10.1016/0550-3213(92)90227-3",
    journal = "Nucl. Phys. B",
    volume = "371",
    pages = "3--31",
    year = "1992"
}

@article{NewMuon:1996uwk,
    author = "Arneodo, M. and others",
    collaboration = "New Muon",
    title = "{Accurate measurement of $F_2^d / F_2^p$ and $R^d - R^p$}",
    eprint = "hep-ex/9611022",
    archivePrefix = "arXiv",
    doi = "10.1016/S0550-3213(96)00673-6",
    journal = "Nucl. Phys. B",
    volume = "487",
    pages = "3--26",
    year = "1997"
}

@article{JeffersonLabHallATritium:2024las,
    author = "Abrams, D. and others",
    collaboration = "Jefferson Lab Hall A Tritium",
    title = "{EMC Effect of Tritium and Helium-3 from the JLab MARATHON Experiment}",
    eprint = "2410.12099",
    archivePrefix = "arXiv",
    primaryClass = "nucl-ex",
    doi = "10.1103/31xz-s84d",
    journal = "Phys. Rev. Lett.",
    volume = "135",
    number = "6",
    pages = "062502",
    year = "2025"
}

@article{BCDMS:1989gtb,
    author = "Benvenuti, A. C. and others",
    collaboration = "BCDMS",
    title = "{A Comparison of the Structure Functions $F_2$ of the Proton and the Neutron From Deep Inelastic Muon Scattering at High $Q^2$}",
    reportNumber = "CERN-EP-89-171",
    doi = "10.1016/0370-2693(90)91232-Z",
    journal = "Phys. Lett. B",
    volume = "237",
    pages = "599--604",
    year = "1990"
}

@article{Alekhin:2022uwc,
	author = "Alekhin, S. I. and Kulagin, S. A. and Petti, R.",
	title = "{Off-shell effects in bound nucleons and parton distributions from $^1$H, $^2$H, $^3$H, and $^3$He data}",
	eprint = "2211.09514",
	archivePrefix = "arXiv",
	primaryClass = "hep-ph",
	doi = "10.1103/PhysRevD.107.L051506",
	journal = "Phys. Rev. D",
	volume = "107",
	number = "5",
	pages = "L051506",
	year = "2023"
}

@article{Alekhin:2022tip,
	author = "Alekhin, S. I. and Kulagin, S. A. and Petti, R.",
	title = "{Nuclear effects in the deuteron and global QCD analyses}",
	eprint = "2203.07333",
	archivePrefix = "arXiv",
	primaryClass = "hep-ph",
	reportNumber = "INR-TH-2022-007",
	doi = "10.1103/PhysRevD.105.114037",
	journal = "Phys. Rev. D",
	volume = "105",
	number = "11",
	pages = "114037",
	year = "2022"
}

@article{Alekhin:2017fpf,
	author         = "Alekhin, S. I. and Kulagin, S. A. and Petti, R.",
	title          = "{Nuclear effects in the deuteron and constraints on the $d/u$ ratio}",
	journal        = "Phys. Rev.",
	volume         = "D96",
	year           = "2017",
	number         = "5",
	pages          = "054005",
	doi            = "10.1103/PhysRevD.96.054005",
	eprint         = "1704.00204",
	archivePrefix  = "arXiv",
	primaryClass   = "nucl-th",
	SLACcitation   = "%%CITATION = ARXIV:1704.00204;%%"
}

@article{Kulagin:2004ie,
      author         = "Kulagin, S. A. and Petti, R.",
      title          = "{Global study of nuclear structure functions}",
      journal        = "Nucl. Phys.",
      volume         = "A765",
      year           = "2006",
      pages          = "126-187",
      doi            = "10.1016/j.nuclphysa.2005.10.011",
      eprint         = "hep-ph/0412425",
      archivePrefix  = "arXiv",
      primaryClass   = "hep-ph",
      SLACcitation   = "%%CITATION = HEP-PH/0412425;%%"
}

@article{Kulagin:2010gd,
	author         = "Kulagin, S. A. and Petti, R.",
	title          = "{Structure functions for light nuclei}",
	journal        = "Phys. Rev.",
	volume         = "C82",
	year           = "2010",
	pages          = "054614",
	doi            = "10.1103/PhysRevC.82.054614",
	eprint         = "1004.3062",
	archivePrefix  = "arXiv",
	primaryClass   = "hep-ph",
	SLACcitation   = "%%CITATION = ARXIV:1004.3062;%%"
}

@article{Kulagin:2014vsa,
	author         = "Kulagin, S. A. and Petti, R.",
	title          = "{Nuclear parton distributions and the Drell-Yan process}",
	journal        = "Phys. Rev.",
	volume         = "C90",
	year           = "2014",
	number         = "4",
	pages          = "045204",
	doi            = "10.1103/PhysRevC.90.045204",
	eprint         = "1405.2529",
	archivePrefix  = "arXiv",
	primaryClass   = "hep-ph",
	reportNumber   = "INR-TH-2014-004, CETUP2014-002",
	SLACcitation   = "%%CITATION = ARXIV:1405.2529;%%"
}

@article{Ru:2016wfx,
    author = "Ru, Peng and Kulagin, S. A. and Petti, R. and Zhang, Ben-Wei",
    title = "{Study of $W^\pm$ and $Z$ boson production in proton-lead collisions at the LHC with Kulagin-Petti nuclear parton distributions}",
    eprint = "1608.06835",
    archivePrefix = "arXiv",
    primaryClass = "nucl-th",
    doi = "10.1103/PhysRevD.94.113013",
    journal = "Phys. Rev. D",
    volume = "94",
    number = "11",
    pages = "113013",
    year = "2016"
}

@article{Akulinichev:1985ij,
	author = "Akulinichev, S. V. and Kulagin, Sergey A. and Vagradov, G. M.",
	title = "The Role of Nuclear Binding in Deep Inelastic Lepton Nucleon Scattering",
	doi = "10.1016/0370-2693(85)90799-3",
	journal = "Phys. Lett. B",
	volume = "158",
	pages = "485--488",
	year = "1985"
}

@article{Kulagin:1989mu,
    author = "Kulagin, S. A.",
    title = "{Deep Inelastic Scattering on Nuclei: Impulse Approximation and Mesonic Corrections}",
    doi = "10.1016/0375-9474(89)90233-9",
    journal = "Nucl. Phys. A",
    volume = "500",
    pages = "653--668",
    year = "1989"
}

@article{Kulagin:1994fz,
	author         = "Kulagin, S. A. and Piller, G. and Weise, W.",
	title          = "{Shadowing, binding and off-shell effects in nuclear deep inelastic scattering}",
	journal        = "Phys. Rev.",
	volume         = "C50",
	year           = "1994",
	pages          = "1154-1169",
	doi            = "10.1103/PhysRevC.50.1154",
	eprint         = "nucl-th/9402015",
	archivePrefix  = "arXiv",
	primaryClass   = "nucl-th",
	reportNumber   = "TPR-94-02, ADP-94-1-T-144",
	SLACcitation   = "%%CITATION = NUCL-TH/9402015;%%"
}

@article{Kulagin:2000yw,
    author = "Kulagin, S. A. and Sidorov, A. V.",
    title = "{Nuclear effects and higher twists in $F_3$ structure function}",
    eprint = "hep-ph/0009150",
    archivePrefix = "arXiv",
    doi = "10.1007/s100500070043",
    journal = "Eur. Phys. J. A",
    volume = "9",
    pages = "261--267",
    year = "2000"
}

@article{Petti:2022bzt,
    author = "Petti, R.",
    title = "{Probing free nucleons with (anti)neutrinos}",
    eprint = "2205.10396",
    archivePrefix = "arXiv",
    primaryClass = "hep-ph",
    doi = "10.1016/j.physletb.2022.137469",
    journal = "Phys. Lett. B",
    volume = "834",
    pages = "137469",
    year = "2022"
}

@article{CLAS:2014jvt,
    author = "Tkachenko, S. and others",
    collaboration = "CLAS",
    title = "{Measurement of the structure function of the nearly free neutron using spectator tagging in inelastic $^2{\mathrm H}(e, e'p)X$ scattering with CLAS}",
    eprint = "1402.2477",
    archivePrefix = "arXiv",
    primaryClass = "nucl-ex",
    reportNumber = "JLAB-PHY-14-1844",
    doi = "10.1103/PhysRevC.89.045206",
    journal = "Phys. Rev. C",
    volume = "89",
    pages = "045206",
    year = "2014",
    note = "[Addendum: Phys.Rev.C 90, 059901 (2014)]"
}

@article{Arrington:2021vuu,
    author = "Arrington, J. and others",
    title = "{Measurement of the EMC effect in light and heavy nuclei}",
    eprint = "2110.08399",
    archivePrefix = "arXiv",
    primaryClass = "nucl-ex",
    doi = "10.1103/PhysRevC.104.065203",
    journal = "Phys. Rev. C",
    volume = "104",
    number = "6",
    pages = "065203",
    year = "2021"
}

\end{document}